\documentclass[%
 aip,
 amsmath,amssymb,
    preprint,%
]{revtex4-1}

\usepackage{graphicx}% Include figure files
\usepackage{verbatim}
\usepackage{dcolumn}% Align table columns on decimal point
\usepackage{bm}% bold math
\usepackage[utf8]{inputenc}
\usepackage[T1]{fontenc}
\usepackage{mathptmx}
\usepackage{etoolbox}
\usepackage{xcolor}

\makeatletter
\def\@email#1#2{%
 \endgroup
 \patchcmd{\titleblock@produce}
  {\frontmatter@RRAPformat}
  {\frontmatter@RRAPformat{\produce@RRAP{*#1\href{mailto:#2}{#2}}}\frontmatter@RRAPformat}
  {}{}
}%
\makeatother
\begin{document}

\preprint{AIP/123-QED}

\title[Spiral Attractors in a Reduced Mean-Field Model of Neuron-Glial Interaction]{Spiral Attractors in a Reduced Mean-Field Model of Neuron-Glial Interaction}
% Force line breaks with \\
\author{S. Olenin}
 \affiliation{Control Theory Department, Lobachevsky University, 603022,  Russia, Nizhny Novgorod, Gagarin ave, 23}%Lines break automatically or can be forced with \\

\author{S. Stasenko}%
\affiliation{ 
Laboratory of advanced methods for high-dimensional data analysis, Lobachevsky University, 603022, Russia, Nizhny Novgorod, Gagarin ave, 23%\\This line break forced with \textbackslash\textbackslash
}%

\author{T. Levanova}
% \homepage{http://www.Second.institution.edu/~Charlie.Author.}
\affiliation{%
Control Theory Department, Lobachevsky University, 603022, Russia, Nizhny Novgorod, Gagarin ave, 23%\\This line break forced% with \\
}%
\email{tatiana.levanova@itmm.unn.ru.}

\date{\today}% It is always \today, today,

\begin{abstract}
This paper investigates various bifurcation scenarios of the appearance of bursting activity in the phenomenological mean-field model of neuron-glial interactions. In particular, we show that the homoclinic spiral attractors in this system can be the source of several types of bursting activity with different properties.
\end{abstract}

\maketitle

\begin{quotation}

It is well known that bursting activity plays an important role in the processes of transmission of neural signals. In terms of population dynamics, macroscopic bursting can be described using a mean-field approach. Mean field theory provides a useful tool for analysis of collective behavior of a large populations of interacting units, allowing to reduce the description of corresponding dynamics to just a few equations. Recently a new phenomenological model was proposed that describes bursting population activity of a big group of excitatory neurons, taking into account short-term synaptic plasticity and the astrocytic modulation of the synaptic dynamics \cite{math11092143}. The purpose of the present study is to investigate various bifurcation scenarios of the appearance of bursting activity in the phenomenological model. We show that the birth of bursting population pattern can be connected both with the cascade of period doubling bifurcations and further development of chaos according to the Shilnikov scenario, which leads to the appearance of a homoclinic attractor containing a homoclinic loop of a saddle-focus equilibrium with the two-dimensional unstable invariant manifold. We also show that the homoclinic spiral attractors observed in the system under study generate several types of bursting activity with different properties. 
\end{quotation}

\section{Introduction}

One of the most interesting and important patterns of population activity is the bursting population activity \cite{chen2022population}, which consists of two stages. The first one includes oscillations with high frequency, repeated at relatively short intervals of time, and the second one is the state of "quiescence". Bursting activity can be observed in many perceptual and behavioral experiments and underlie both normal physiological \cite{meister1991synchronous,krahe2004burst,salinas2001correlated,axmacher2006memory} and pathophysiological processes (e.g., epilepsy \cite{avoli1987bursting,hofer2022bursting,stasenko2023loss}). Among the various informational and physiological processes related to bursting activity the following ones can be mentioned: restoration of synaptic transmission after its disruption \cite{lisman1997bursts}, neurotransmitter release \cite{lisman1997bursts}, formation of long-term potentiation \cite{lisman1997bursts}, selectivity of synaptic communication \cite{izhikevich2003bursts}, sensory inputs and increasing the information capacity \cite{gabbiani1996stimulus,oswald2004parallel,lesica2004encoding,reinagel1999encoding,segev2004hidden} and many others \cite{hulata2004self,stasenko2023information,stasenko2023dynamic}.

The mechanisms underlying the appearance of bursting population activity are not yet fully understood \cite{gast2020mean}. To date, most computational models deal with the study of single cell dynamics, see, e.g. \cite{izhikevich2003bursts,guckenheimer1997bifurcation}. 
To reproduce bursting activity at the population level one can use implicit properties of interacting excitatory and inhibitory populations \cite{kudela2003changing,zeldenrust2018neural} or explicitly add burst-provoking factors (e.g. action of a neuromodulator \cite{marder2002cellular}, feedforward inhibition \cite{zeldenrust2013modulation}, spike-frequency adaptation \cite{vreeswijk2001patterns} or their combinations \cite{markram1996redistribution, masquelier2013network,rozhnova2021bifurcation,stasenko2023bursting}). Another valuable features that should be taken into account when creating such models are astrocytic modulation of neuronal activity and short-term synaptic plasticity \cite{barabash2021stsp,barabash2023rhythmogenesis}. Adding these factors to the model can lead to rich and nontrivial dynamics \cite{stasenko2020quasi,stasenko20223d}. In particular, in \cite{cortes2013short} it was shown that the addition of short-term synaptic plasticity to  the Tsodyks-Markram model leads to the emergence of chaotic bursting related to spiral chaos, which occurs according to the Shilnikov scenario \cite{gonchenko2022leonid}.

Shilnikov given a phenomenological description of his scenario in the paper \cite{shilnikov1984bifurcation} for the case of one-parameter families of multidimensional systems. This scenario is extremely simple (by modulo of certain intermediate details that themselves can be very complicated), and therefore, it should come as no surprise that it is often seen in many models. In the three-dimensional case, the essence of the Shilnikov scenario is as follows.

At the beginning, the attractor of a system is a stable equilibrium. Then it loses stability under a supercritical (soft) Andronov-Hopf bifurcation: the equilibrium itself becomes a saddle-focus of type (1,2), i.e. with one-dimensional stable and two-dimensional unstable invariant manifolds, and a stable limit cycle is born, which becomes the attractor. With further changes in parameter, this cycle becomes focal, i.e. it takes complex conjugate multipliers, and the so-called  Shilnikov funnel is created, into which all trajectories from the absorbing region are drawn, with the exception of one stable separatrix of the saddle-focus. This funnel persists even when the limit cycle loses its stability and the attractor becomes strange. Moreover, the Shilnikov attractor can be formed here, when  a homoclinic loop of the saddle-focus arises (one of its stable separatrices falls on the two-dimensional unstable manifold). In this case, the famous Shilnikov criterion of chaos \cite{shilnikov1965case} is certainly satisfied when the divergence at the saddle-focus is negative. 

It is worth noting that Shilnikov chaos has been observed in various systems, both in experiments and in models. It has been experimentally detected in various systems, including laser systems \cite{pisarchik2000discrete,pisarchik2001synchronization,pisarchik2001theoretical, arecchi1987laser,viktorov1995shil}, optically bistable devices \cite{farjas1996equivalent}, electronic circuits \cite{kingni2013dissipative}, memristive circuits \cite{njitacke2016coexistence}, and numerous other systems \cite{dangoisse1988shilnikov,hennequin1989characterization,de1989instabilities}. Additionally, numerical investigations have revealed the appearance of Shilnikov chaos in mathematical models of diverse scientific fields, such as biology and medicine \cite{griffith2009vasomotion,cortes2013short,korotkov2019effects,shilnikov2008methods}, ecology \cite{bakhanova2018spiral,barrio2011global,rosenzweig1973exploitation,hastings1991chaos}, chemistry \cite{gaspard1983can}, economics \cite{barnett2022shilnikov,bella2017shilnikov}, engineering \cite{wei2010shilnikov,yao2005multipulse}, and many others, see, e.g., papers \cite{borisov2016spiral,bakhanova2020homoclinic,bakhanova2023shilnikov,malykh2020homoclinic,karatetskaia2021shilnikov} and references therein.

One of the main features of Shilnikov chaos is that one can distinguish in it  two rather different stages: the stage during which the trajectory passes near the equilibrium point (the stage of small amplitude), and the stage, when it passes along a global piece of the homoclinic loop (the stage of large amplitude). These stages are the source of specific bursting patterns (so-called homoclinic burstings) that differ from typical patterns of bursting activity when oscillations with high frequencies alternate with stages of quiescence (non-homoclinic burstings). In \cite{bakhanova2018spiral}, using the Rosenzweig-MacArthur system as an example, it was shown that homoclinic attractors, which appear due to Shilnikov bifurcation scenario, can be a source of such types of bursting activity.

In this paper, we focus on similar scenarios of the appearance of different types of regular and chaotic bursting population activity in a recently proposed mean-field model of neuron-glial interaction \cite{math11092143}. %In particular, we explore in detail the emergence of chaotic dynamics associated with spiral attractors in a recently proposed 3-dimensional phenomenological model , which describes the synchronous population dynamics of a groups of neurons. 
This model is a simplification of the previously proposed 4-dimensional model of population activity \cite{barabash2021stsp,barabash2023rhythmogenesis}, which, in its turn, is based on the Tsodyks-Markram mean-field model \cite{mongillo2008synaptic} and takes into account the main features of neuron-glial interaction \cite{lazarevich2017synaptic}. We show that mixed-mode oscillations arising from a Shilnikov's homoclinic loop, together with a slow-fast behavior, lead to the appearance of homoclinic bursting population activity in the system.

The paper is organized as follows. In Section \ref{sec:model} we introduce and briefly describe the reduced model of population activity. We present the results of two-parameter bifurcation analysis in Section \ref{sec:bifurc}. We describe the onset of spiral chaos according to Shilnikov scenario in Section \ref{sec:shilnikov}. In Section \ref{sec:results} we compare different types of bursting activity based on its features and mathematical images. In Section \ref{sec:discussion} we discuss our results before draw our conclusions in Section \ref{sec:conclusion}.

\section{The model}
\label{sec:model}

To investigate mechanisms underlying population bursting activity in the presence of astrocytic influence, in our previous paper \cite{math11092143} we proposed a new reduced mean-field model of neuron-glial interactions. This model is a combination of two models, the Wilson-Cowan model \cite{wilson1972excitatory} describing the dynamics of populations of excitatory and inhibitory neurons, and the Tsodyks-Markram model \cite{mongillo2008synaptic,tsodyks1998neural} of short-term synaptic plasticity, to which we added the equation of neuron-glial interaction, phenomenologically described using the mean- field approach \cite{stasenko2020quasi,gordleeva2012bi}. For simplicity, we focus only on the population of excitatory neurons, as done in \cite{wilson1972excitatory}. 

The model describes the synchronous dynamics of a population of excitatory neurons (e.g., pyramidal neurons) with a constant inhibitory input from another population of neurons (e.g., interneurons) in the presence of astrocytic influence. Mathematically, the reduced model can be written in the form of the following three-dimensional system of ODEs:
\begin{equation}
\left\{\begin{array}{l}
    \tau \dot E = -E + \alpha \ln{(1+\exp{\frac{JU(y)xE+I_0}{\alpha}})}\\
    \dot x = \frac{1 - x}{\tau_D} - U(y)xE\\
    \dot y = \frac{-y}{\tau_y}+\beta \sigma_y(x)\\
\end{array}
\right.\label{eq:TM}
\end{equation}

Here the phase variables $E,x$ and $y$ are as follows: $E$ is the value of synchronous activity of population of excitatory neurons, $x$ and $y$ are, respectively, probabilities of neurotransmitter  and gliotransmitter release. Note that variable $E$ is not a membrane potential, but only reflects the rate of population activity of a large population of neurons. System \eqref{eq:TM} depends also on a number of parameters that are certain characteristics of phenomenological functions of phase variables used for the modelling of synaptic processes. So, the parameter $I_0$ is a constant that defines the inhibitory input from the population of interneurons applied to the population of excitatory neurons. The parameter $\alpha$ determines the threshold for increasing the synchronous activity of the population of excitatory neurons. Here the parameters $\tau$, $\tau_D$ and $\tau_y$ are characteristic time constants for population activity, synaptic depression and gliotransmitter relaxation, respectively. 

Function $U(y)$ determines the probability of neurotransmitter release, while the combination $JU(y)xE$ introduces a positive feedback:
\begin{equation}
   U(y) = U_{0}  + \frac{\Delta U_{0}}{1 + e^{-50(y-y_{thr})}},
   \label{eq:astro_influence}
\end{equation}
where the parameter $U_{0}$ determines an averaged probability of neurotransmitter release, the parameter $\Delta U_{0}$ corresponds to the change in $U_0$ induced by the astrocytic influence, the parameter $y_{thr}$ is the activation threshold for the gliotransmitter  that sets a level of gliotransmission required to manifest an effect of neurotransmitter release.

Function $\sigma_y(x)$ is an activation function for astrocyte that depends on the state of neuron:
\begin{equation}
\sigma_y(x) = \frac{1}{1 + e^{-20(x-x_{thr})}},\label{eq:astro}
\end{equation} 
with activation threshold $x_{thr}$ determines the level of neural activity at which the gliotransmitter is released by astrocyte. In system \eqref{eq:TM} parameter $\beta$ preceding the function $\sigma_y(x)$ sets the released part of the gliotransmitter.

The interaction between neurons and astrocytes can be simplistically described as follows. Being activated, excitatory neurons release the neurotransmitter $x$, which, in turn, activates astrocytes via a cascade of biochemical reactions. As a result, a gliotransmitter $y$ is released. The gliotransmitter, in its turn, can change the probability of neurotransmitter release. Thus, a feedback loop is formed \cite{araque1999tripartite}. This phenomenon is depicted in the model in such a way that the function $U$ (activation for neuron) depends on $y$, and the function $\sigma$ (activation for astrocyte) depends on $x$.

It should be noted that the proposed model does not incorporate the mechanism of synaptic facilitation. The model includes short-term synaptic plasticity only in the form of synaptic depression. The regulation of different patterns of population activity in the model is solely controlled by the dynamics of the astrocytes through astrocytic facilitation of synaptic transmission. Our model has all the limitations inherent in a Wilson-Cowan type model.

We choose $I_0$ and $U_0$ as control parameters of the model as it was done in \cite{math11092143}. All other parameters were fixed as follows: $\Delta U_0 =0.305$, $\tau_{D} = 0.07993$, $\tau_{y} = 3.3$, $\tau= 0.013$,  $\beta = 0.3$, $x_{thr}=0.75$, $y_{thr}=0.4$, $J=3.07$, $\alpha = 1.58$.

\section{Two-parameter bifurcation analysis}
\label{sec:bifurc}

In our recent paper \cite{math11092143} it was shown that the dynamics of system \eqref{eq:TM} can be chaotic. In this paper we go beyond the above-mentioned studies and explore peculiarities of chaotic bursting population activity in system \eqref{eq:TM}. 

\begin{figure}[tbh]
	\centering
    \includegraphics[width=\textwidth]{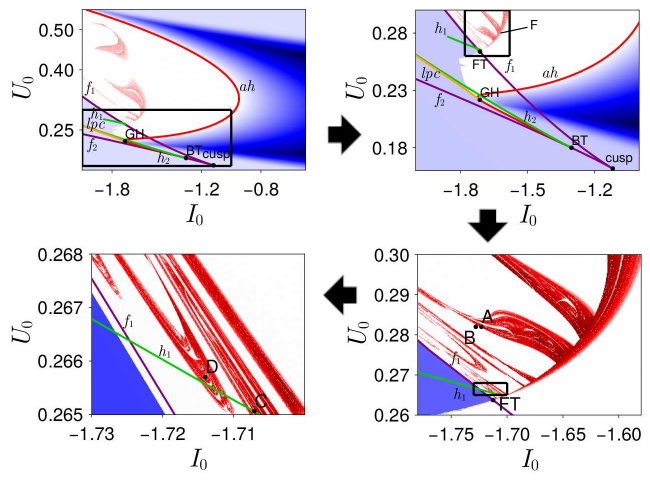}
	\caption{Chart of MLE $\lambda$ on the parameter plane ($I_0$, $U_0$) and its enlarged fragments. Red marker corresponds to the regions of chaotic dynamics ($\lambda>0$), blue marker -- to the regions of quiescence ($\lambda<0$), white marker -- to the regions with periodic dynamics ($\lambda=0$). Here $ah$ label marks Andronov-Hopf bifurcation, $f_1$ and $f_2$ -- fold bifurcations, $h_1$ and $h_2$ -- homoclinic bifurcations. Point $BT$ marks the Bogdanov-Takens bifurcation, point $GH$ -- generalized Hopf bifurcation, $cusp$ -- cusp point. Letters $A$ and $B$ in right lower panel and $C$ and $D$ in left lower panel mark combinations of parameters $I_0$ and $U_0$ that correspond to different patterns of bursting population activity. See detailed description in the text.}
	\label{fig:LLE_map}
\end{figure}

In order to do this, we conducted a two-parameter bifurcation analysis of system \eqref{eq:TM}. First of all, we constructed the chart of maximal Lyapunov exponent (MLE) $\lambda$ on the plane of control parameters $I_0$ and  $U_0$, see fig. \ref{fig:LLE_map}. This chart consists of different colored regions whose colors mean the following. The red colored regions correspond to those parameter values at which system \eqref{eq:TM} has chaotic dynamics (strange attractor, $\lambda>0$). Regions with shades of blue correspond to those parameter values at which system \eqref{eq:TM} has a stable equilibrium (a "quiescence" state, $\lambda<0$). Finally, white regions  correspond to those parameter values at which system \eqref{eq:TM} has a stable limit cycle (a periodic oscillations, $\lambda=0$ and other Lyapunov exponents are negative).

As can be seen from fig. \ref{fig:LLE_map}, chaotic regions include stability windows, i.e. regions in the parameter space which correspond to the emergence of stable periodic orbits (here, multi-turn limit cycles). Such windows of stability are typical for chaotic non-hyperbolic systems and have universal form called 'saddle area' (or 'squid'), see \cite{gonchenko2022universal} for details.

Also, we performed a numerical bifurcation analysis of system \eqref{eq:TM}. Namely, using Bifurcationkit package \cite{veltz} and Matcont software \cite{matcont}, we plotted in the MLE chart bifurcation curves  both for equilibria of system \eqref{eq:TM} and for homolinic loops to these equilibria of saddle and saddle-focus type.

Three  of these curves $ah$, $f_1$ and $f_2$ relate to bifurcations of equilibria. 

Curves $f_1$ and $f_2$ relate to the fold bifurcations. Note that in the region between the curves $f_1$ and $f_2$ (let us call this region $Q_3$), which converge at the $cusp$ point, there are three equilibria $O_1, O_2$ and $O_3$, and outside this region there is only one. The equilibria $O_2$ and $O_3$ merge and disappear as a result of fold bifurcation when passing through the curve $f_1$. Here this bifurcation is always of a saddle-node type: upon entering the region $Q_3$, a stable equilibrium $O_3$ and a saddle $O_2$ of type (2,1) are born. On the curve $f_2$, where the equilibria $O_1$ and $O_2$ merge, there is a bifurcation point $BT$ of codimension 2, the so-called Bogdanov-Takens point, that divides the curve $f_2$ into two segments, from the point $BT$ to the $cusp$ point, where the equilibrium $O_{12}$ is a saddle-node, and from the point $BT$ to the left, where $O_{12}$ is a saddle-saddle.

The red curve $ah$ corresponds to an Andronov-Hopf bifurcation of the equilibrium $O_1$. On this curve there is the bifurcation point $GH$ of codimension 2, the so-called generalized Hopf (or Bautin \cite{Kuz}) bifurcation point, which divides the curve $ah$ into two segments, above the point $GH$, where the bifurcation is supercritical, and below the point $GH$, where the bifurcation is subcritical. During the supercritical Andronov-Hopf bifurcation, the stable point $O_1$ becomes a saddle-focus (1,2) and a stable limit cycle is born. During the subcritical Andronov-Hopf bifurcation, the stable point $O_1$ is surrounded by a saddle limit cycle, which disappears as a result of sticking into the point $O_1$, which becomes a saddle-focus (1,2).

It is worth to note some features of dynamics of system \eqref{eq:TM} for parameter values from the region $Q_3$. Firstly, there is a phenomenon of multistability for some parts of this region. Thus, near the $cusp$ point the equilibria $O_1$ and $O_3$ are attractors; in some subregion above the curve $ah$, attractors are $O_3$ and the limit cycle $C_1$ that was born from $O_1$. Also here one can, in principle, observe such types of multistability when the attractor $O_3$ coexists with attractors generated by bifurcations of the cycle $C_1$ (including chaotic ones after an infinite cascade of bifurcations). In addition, it should also be noted that when leaving the region $Q_3$ through a part of the curve $f_1$ above the $FP$ point (Feigenbaum point), one can observe the scenario of transition to chaos via  intermittency \cite{PomMan80}.

There are also several bifurcation points of codimension two presented in fig. \ref{fig:LLE_map}. The point $GH$ on the curve $ah$ marks a generalized Hopf bifurcation. The higher branch of $ah$ curve corresponds to supercritical Andronov-Hopf bifurcation, while the lower branch -- to subcritical Andronov-Hopf bifurcation. At the bifurcation point $GH$ the equilibrium $O_1$ has a pair of purely imaginary eigenvalues and a zero first Lyapunov coefficient. Point $BT$ on the curve $f_2$ marks the Bogdanov-Takens bifurcation. At $BT$ bifurcation point, three bifurcation curves intersect: the Andronov-Hopf bifurcation curve $ah$, the fold bifurcation curve $f_2$ and the homoclinic bifurcation curve $h_2$. At the bifurcation point $BT$ the equilibrium $O_1$ has the double zero eigenvalue. In the $cusp$ bifurcation point, where equilibrium state has zero eigenvalue and the quadratic coefficient for the saddle-node bifurcation vanishes, two fold bifurcation curves $f_1$ and $f_2$ converge. 

\begin{figure}[tbh]
    \centering
    (a) \includegraphics[width = 6.0cm,  height = 5.5cm]{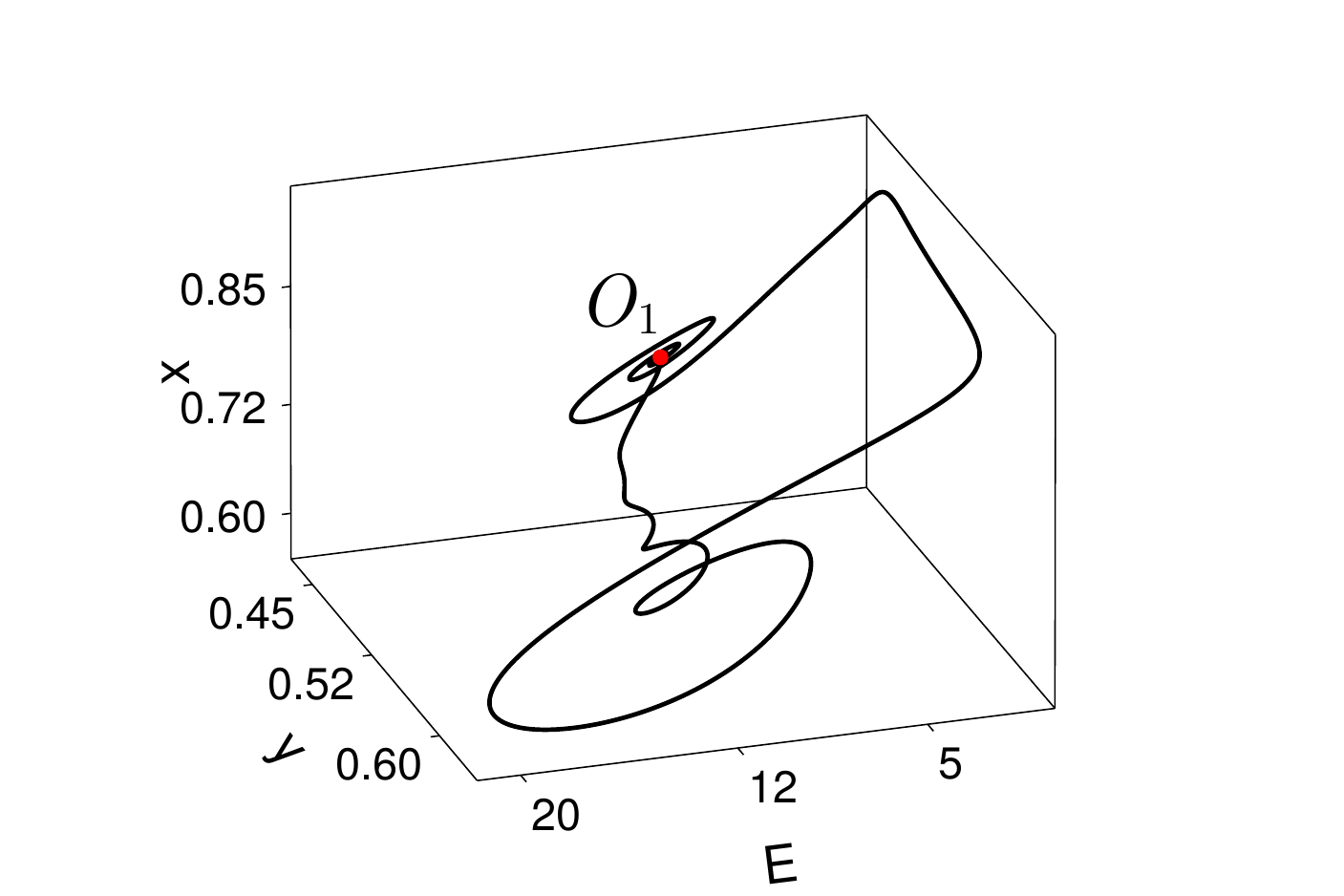}
    (b) \includegraphics[width = 6.0cm,  height = 5.5cm]{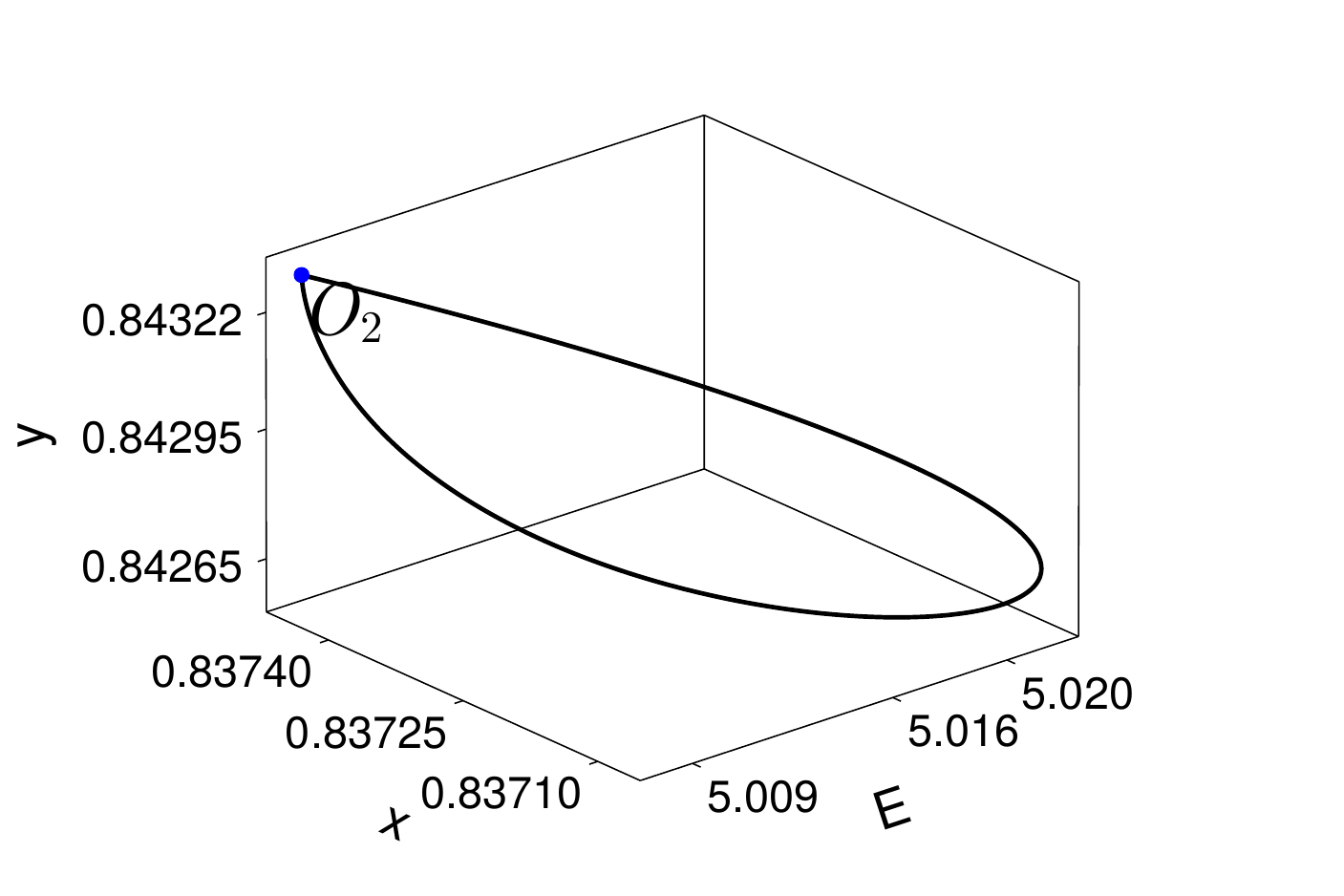}
	\caption{Homoclinic loop to (a) saddle-focus $O_1$ for $I_0 \approx -1.7187$, $U_0 \approx 0.2659$, (b) saddle $O_2$ for $I_0 \approx -1.3032$, $U_0 \approx 0.1799$.}
	\label{fig:homoclinic_to_saddle}
\end{figure}

Of particular interest for further research are two green curves $h_1$ and $h_2$, which mark homoclinic bifurcations. On the curve $h_1$ a homoclinic loop to saddle-focus $O_1$ appears (see fig. \ref{fig:homoclinic_to_saddle}(a)), while on the curve $h_2$ homoclinic loop to a saddle $O_2$ arises (see fig. \ref{fig:homoclinic_to_saddle}(b)). 

Now let us study in detail the parameter region near $h_1$ homoclinic curve, since it can be a source of spiral homoclinic attractors born due to Shilnikov scenario and, therefore, a source of different patterns of bursting activity associated with it.

\section{Shilnikov scenario and homoclinic attractors}
\label{sec:shilnikov}

\begin{figure}[tbh]
    \centering
    (a)\includegraphics[width=4.5cm,  height = 4.5cm]{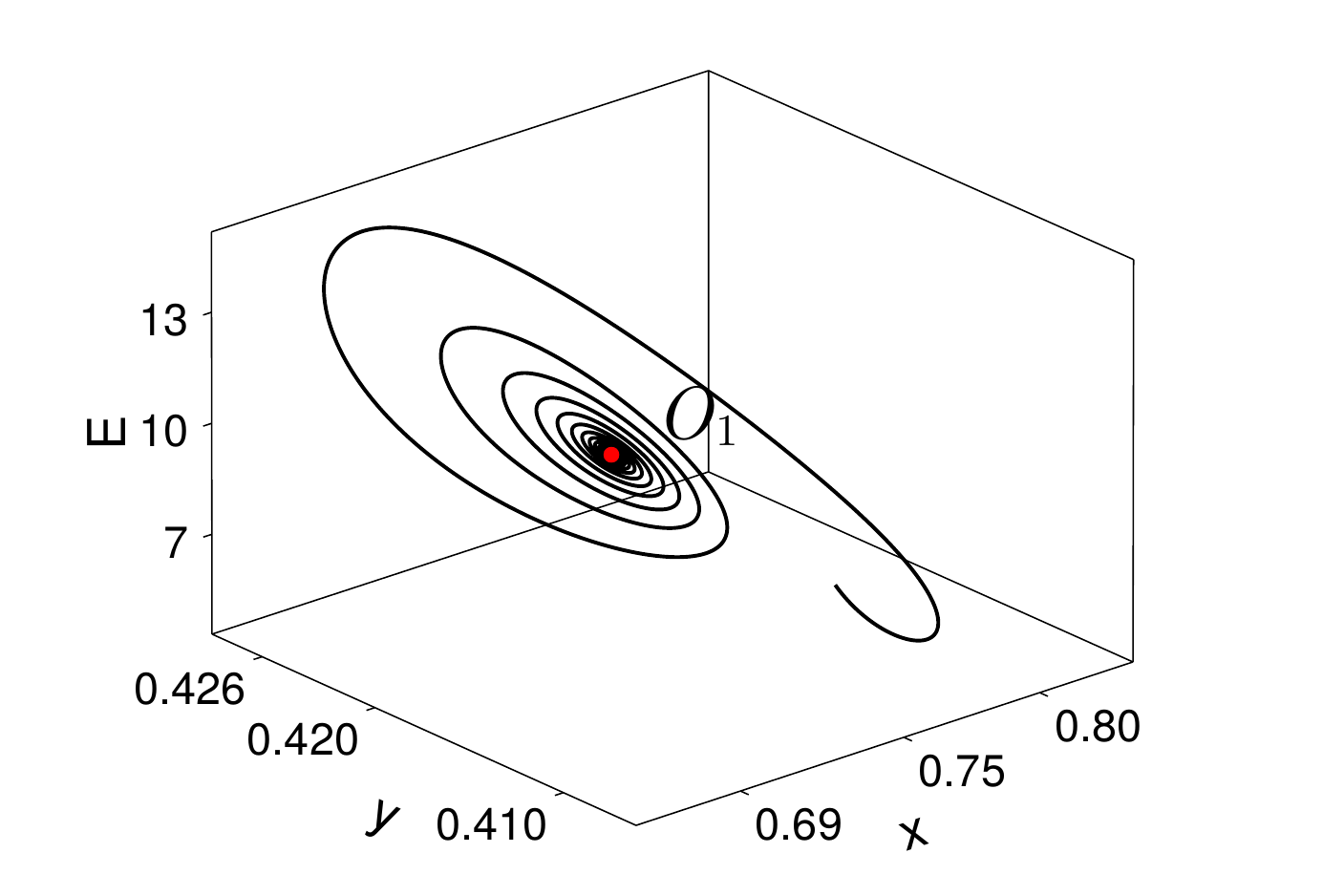}
    (b)\includegraphics[width=4.5cm,  height = 4.5cm]{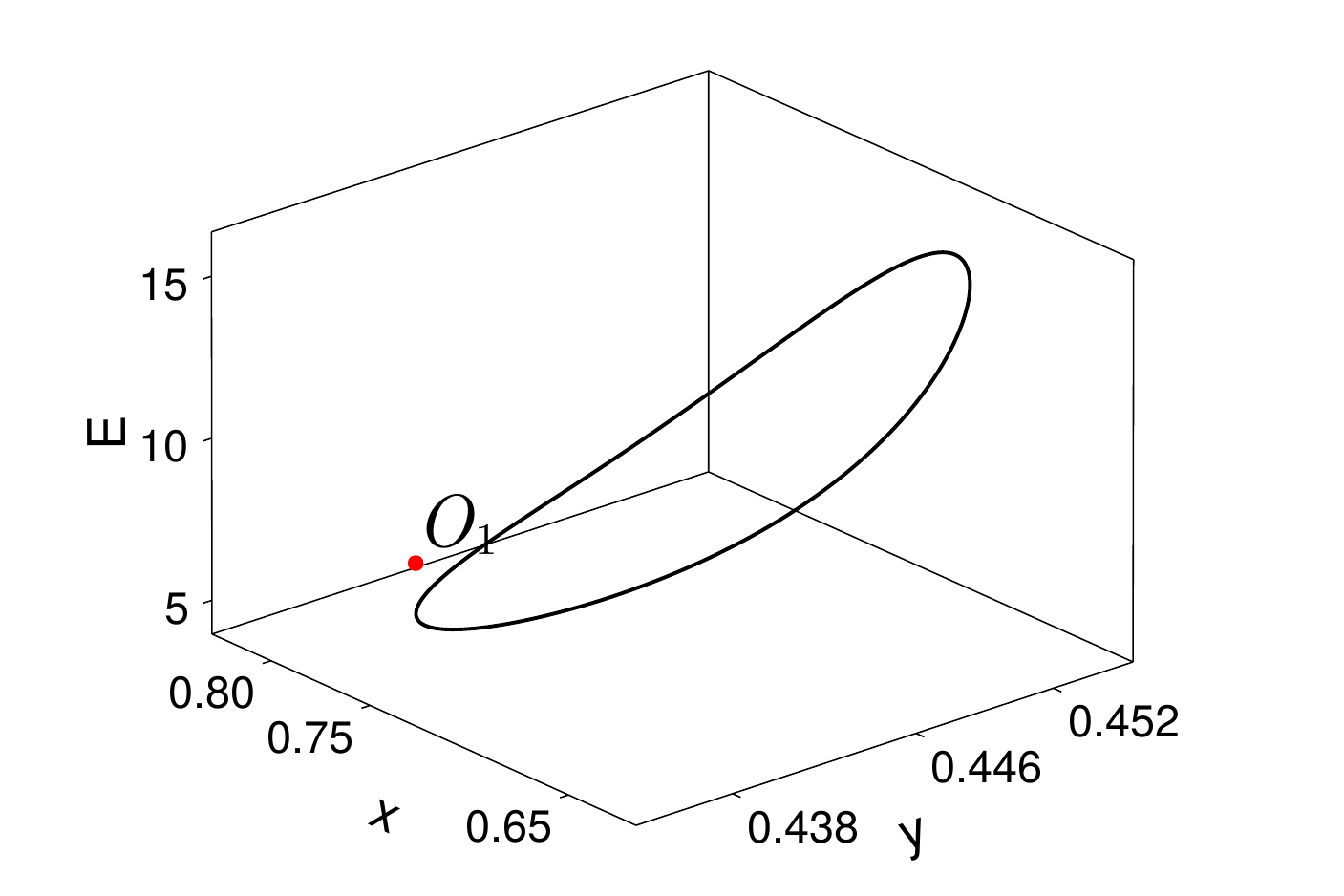}
    (c)\includegraphics[width=4.5cm,  height = 4.5cm]{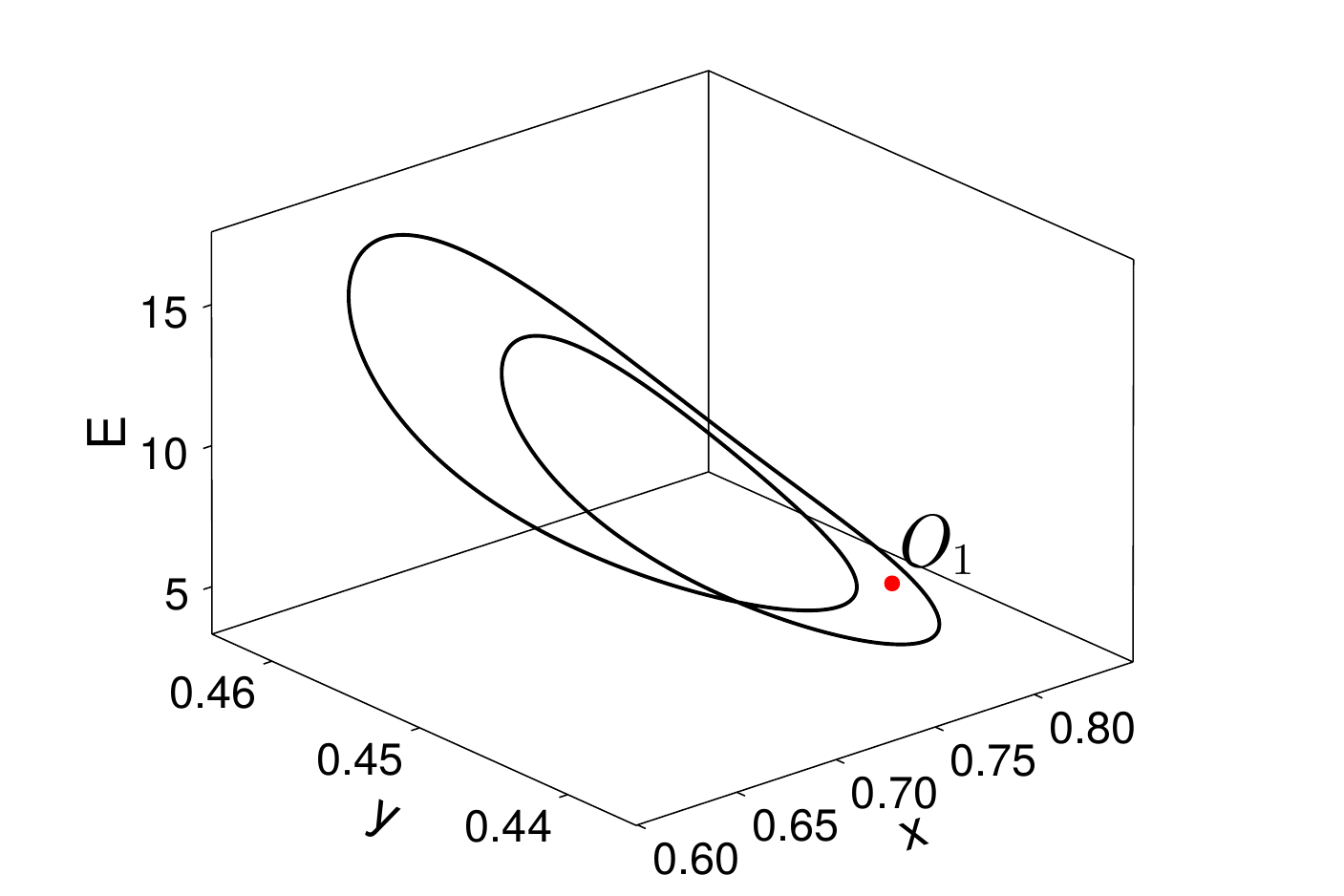}\\
    (d)\includegraphics[width=4.5cm,  height = 4.5cm]{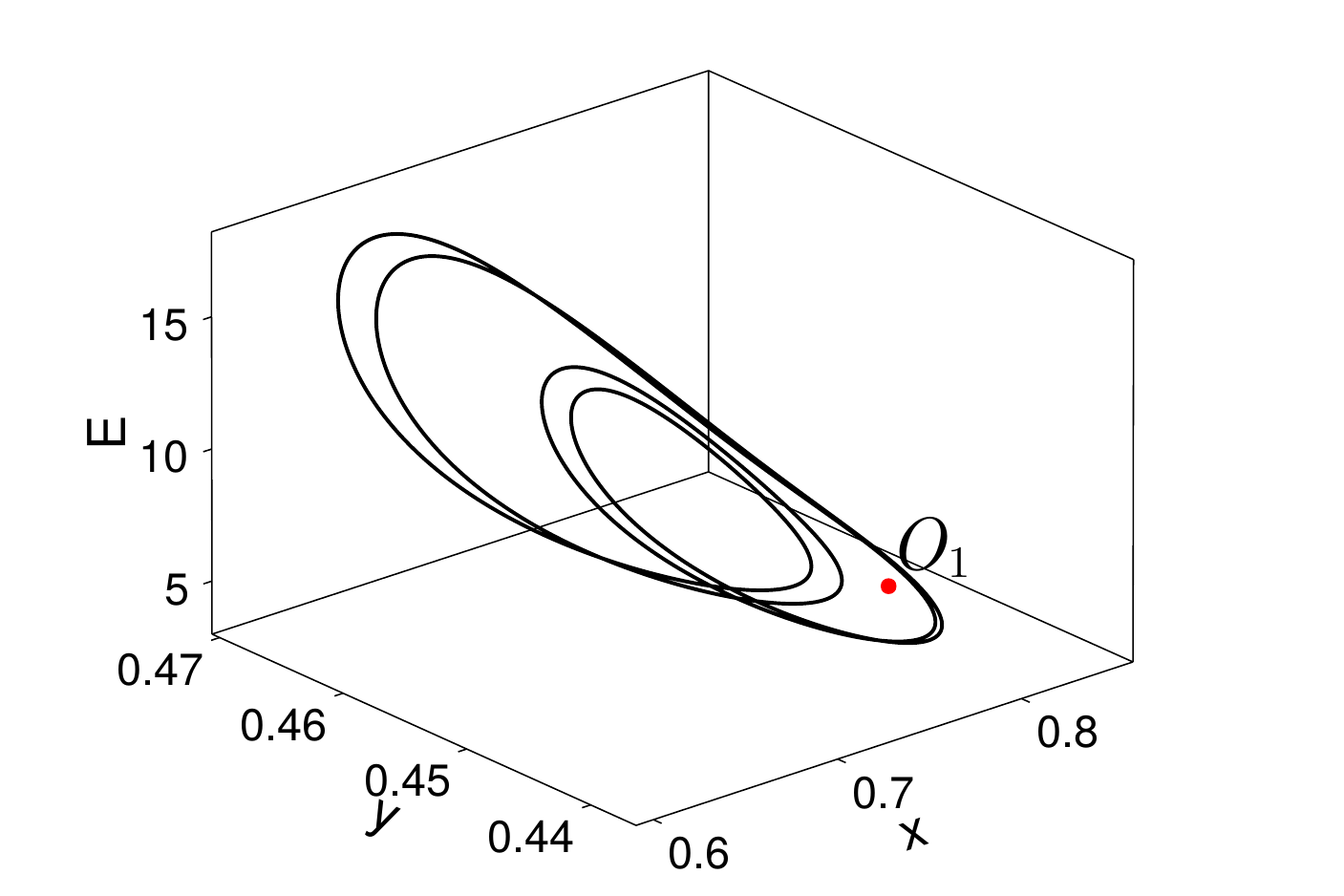}
    (e)\includegraphics[width=4.5cm,  height = 4.5cm]{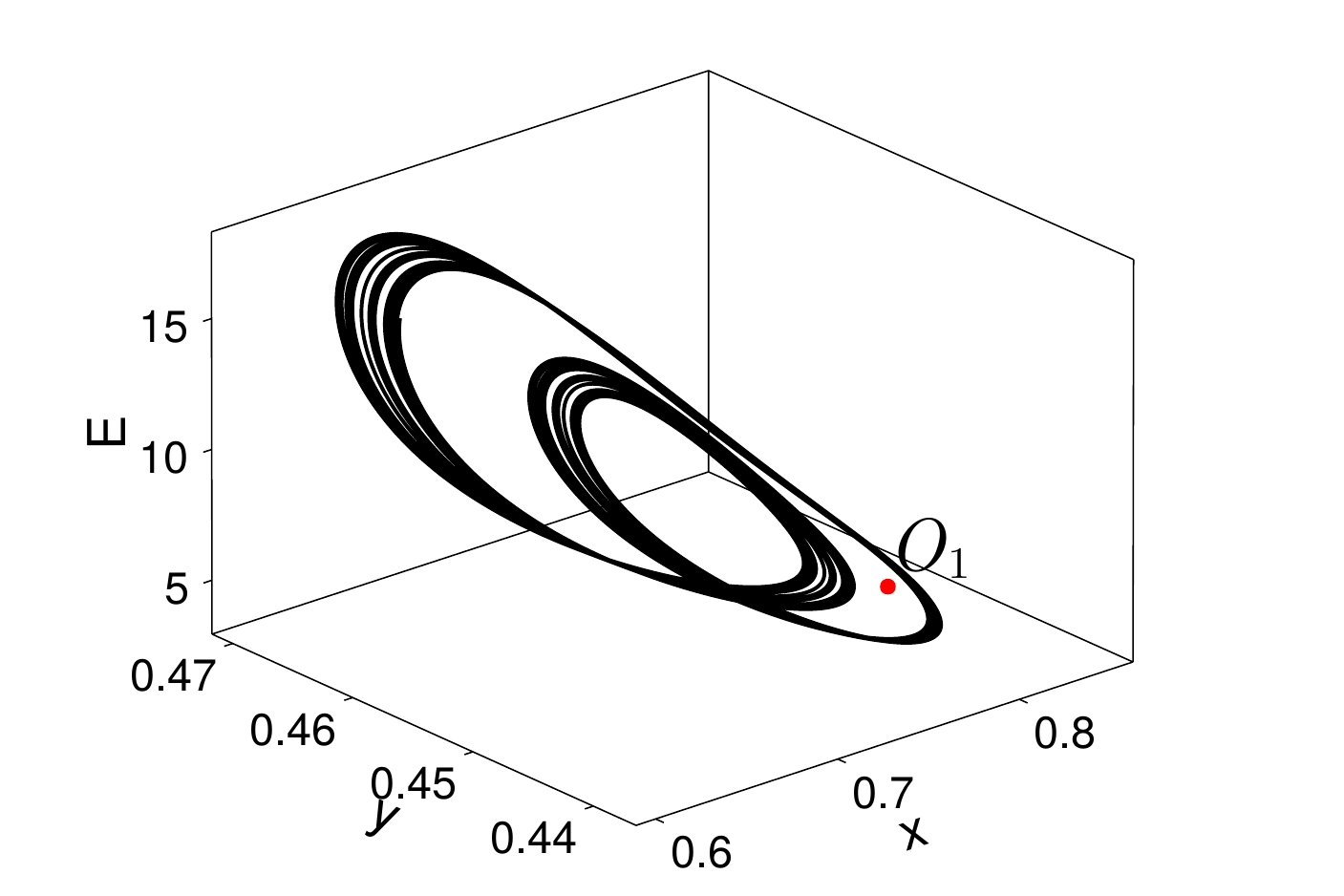}
    (f)\includegraphics[width=4.5cm,  height = 4.5cm]{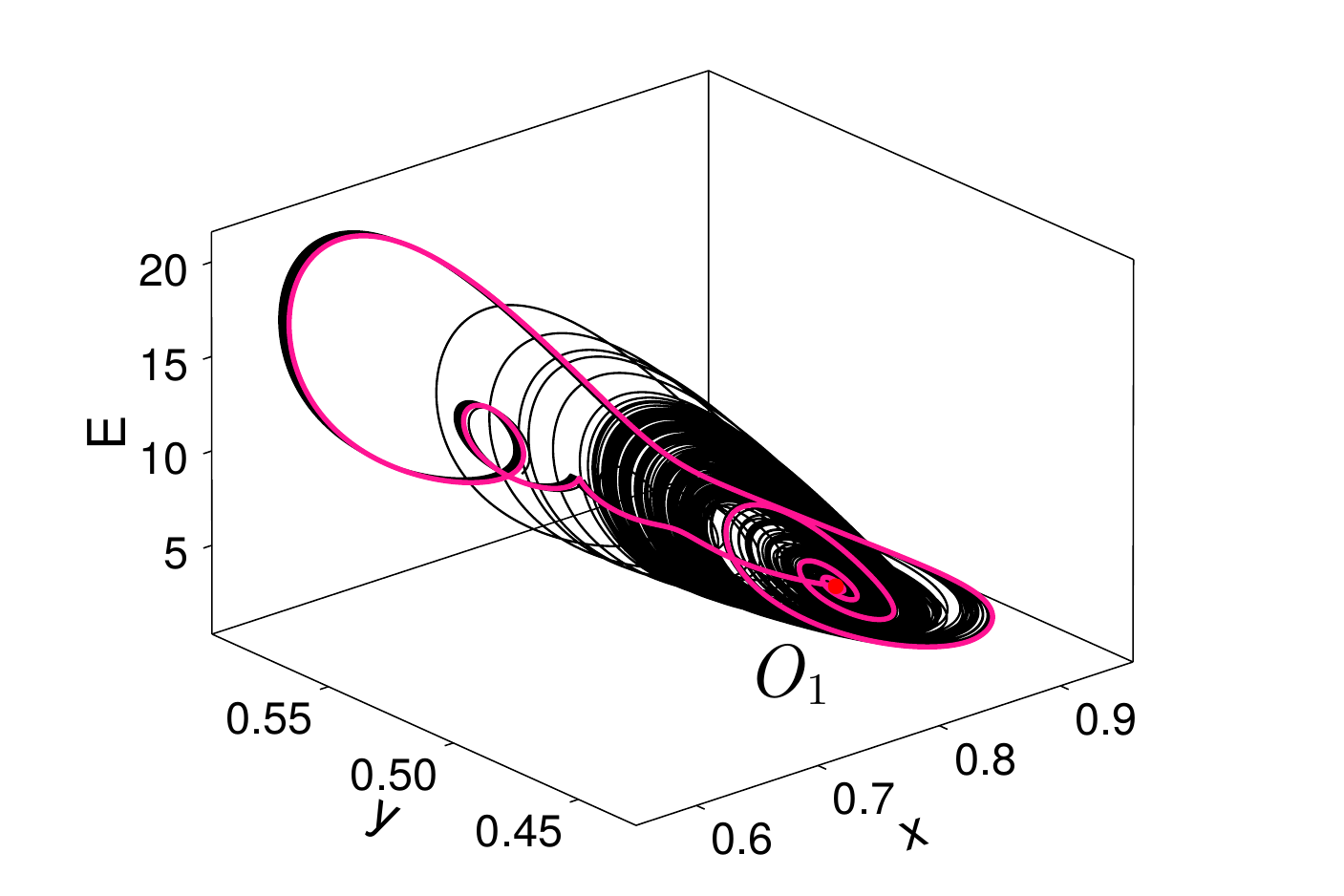}   
	\caption{Shilnikov's scenario in system \eqref{eq:TM}: (a) stable equilibrium $O_1$, $I_0 = -0.9$; (b) stable limit cycle, $I_0 = -1.6$; (c) two-period stable limit cycle, $I_0 = -1.66$; (d) 4-period stable limit cycle, $I_0 = -1.695$; (e) attractor after a cascade of period doubling bifurcations, $I_0 = -1.6972$; and (f) a homoclinic spiral attractor, $I_0 = -1.706$, the pink curve corresponds to the homoclinic trajectory, the dark grey curves -- to the trajectories on the attractor. In all cases $U_0 = 0.265$. In cases (b)-(f) saddle-focus $SF(1,2)$ is depicted near attractor.}
	\label{fig:Shilnikov_scenario}
\end{figure}

Let us describe the main stages of Shilnikov's scenario that is observed in system \eqref{eq:TM} for fixed $U_0 = 0.265$ and decreasing $I_0$, see fig. \ref{fig:Shilnikov_scenario}. For $I_0 > I_{ah} \approx -1.1065337$, there is an asymptotically stable equilibrium state $O_1$ (fig. \ref{fig:Shilnikov_scenario}(a)). For $I_0 = I_{ah}$ it undergoes a supercritical Andronov-Hopf bifurcation, as a result of which the equilibrium $O_1$ becomes a saddle-focus $(1,2)$ and a stable limit cycle $C_1$ is born (fig. \ref{fig:Shilnikov_scenario}(b)). Then, for further decrease in the value of $I_0$ this limit cycle becomes focal and the two-dimensional unstable manifold $W^u (O_1)$ begins to wrap around it, forming a funnel-like configuration. The boundary of this funnel consists of a saddle-focus equilibrium and its two-dimensional unstable invariant manifold, which has the shape of a bowl, the edges of which are bent inside the funnel. Besides, all trajectories from the absorbing region are drawn into the funnel,  except for one stable separatrix of $O_1$ that tends to $O_1$ from the outer side of the funnel. With further varying $I_0$, the limit cycle $C_1$ undergoes a cascade of period-doubling bifurcations (fig. \ref{fig:Shilnikov_scenario}(c)-(d)), while the size of the funnel grows. As a result, a strange attractor of Feigenbaum type appears (fig. \ref{fig:Shilnikov_scenario}(e)).  Finally, for $I_0 \approx -1.7062$ a homoclinic loop to saddle-focus $O_1$ appears. In fig. \ref{fig:Shilnikov_scenario}(f) one can see a strange attractor containing this loop. Strange attractors of such type, containing either an equilibrium state in the case of flows or a fixed point in the case of mappings, are generally called homoclinic attractors \cite{GGS12,GGKT14}. In this particular case, the attractor has name the Shilnikov attractor \cite{GGKKB19}. 

The structure of the set of trajectories of spiral attractors can be quite diverse. The basic elements of the geometry of such attractors form mainly those trajectories that pass near a global piece of the unstable manifold $W^u(O1)$, which, in turn, is the boundary of the Shilnikov funnel. These trajectories can make one, two, three or more turns before returning to the vicinity of the saddle-focus. In papers devoted to the study of the topology of attractors, e.g., in \cite{rossler1979continuous, argoul1987experimental, letellier1995unstable}, such attractors are often called screw, funnel and multifunnel attractors. The type of spiral attractor (connected with the a number of turns) affects the number of oscillations in the additional high-amplitude stage of bursting \cite{bakhanova2018spiral}.

Note that spiral attractors with different topology appear in models of single neurons \cite{hindmarsh1984model,shilnikov2008methods}, as well as in the classical mean-filed models \cite{cortes2013short}. Moreover, in all these cases the onset of chaos occurs due to the Shilnikov scenario. 

\section{Different types of bursting activity in a reduced model of neuron-glial interaction}
\label{sec:results}

System \eqref{eq:TM} belongs to the class of slow-fast systems ($E$ is the fast variable, while $x$ and $y$ are slow ones), which also have spiral attractors. It is therefore not surprising that system \eqref{eq:TM} can exhibit various types of bursting population activity. Since spiral attractors belong to the class of quasiattractors, they either contain stable periodic orbits (as a rule, of big periods and with thin absorbing domains)  or such orbits arise under arbitrarily small perturbations. These periodic orbits repeat a global behavior of homoclinic loops in the attractors, so one can often observe near a chaotic spiral attractor its "a regular version" (a multi-round stable limit cycle existing  for parameter values from the corresponding windows of stability). In contrast to spiral attractors, that exist on certain bifurcation line, stable limit cycles persist under small perturbations and exist for some open regions of the parameter space.

\begin{figure}[tbh]
	\centering
    (a)\includegraphics[width=9.0cm,  height = 9.0cm]{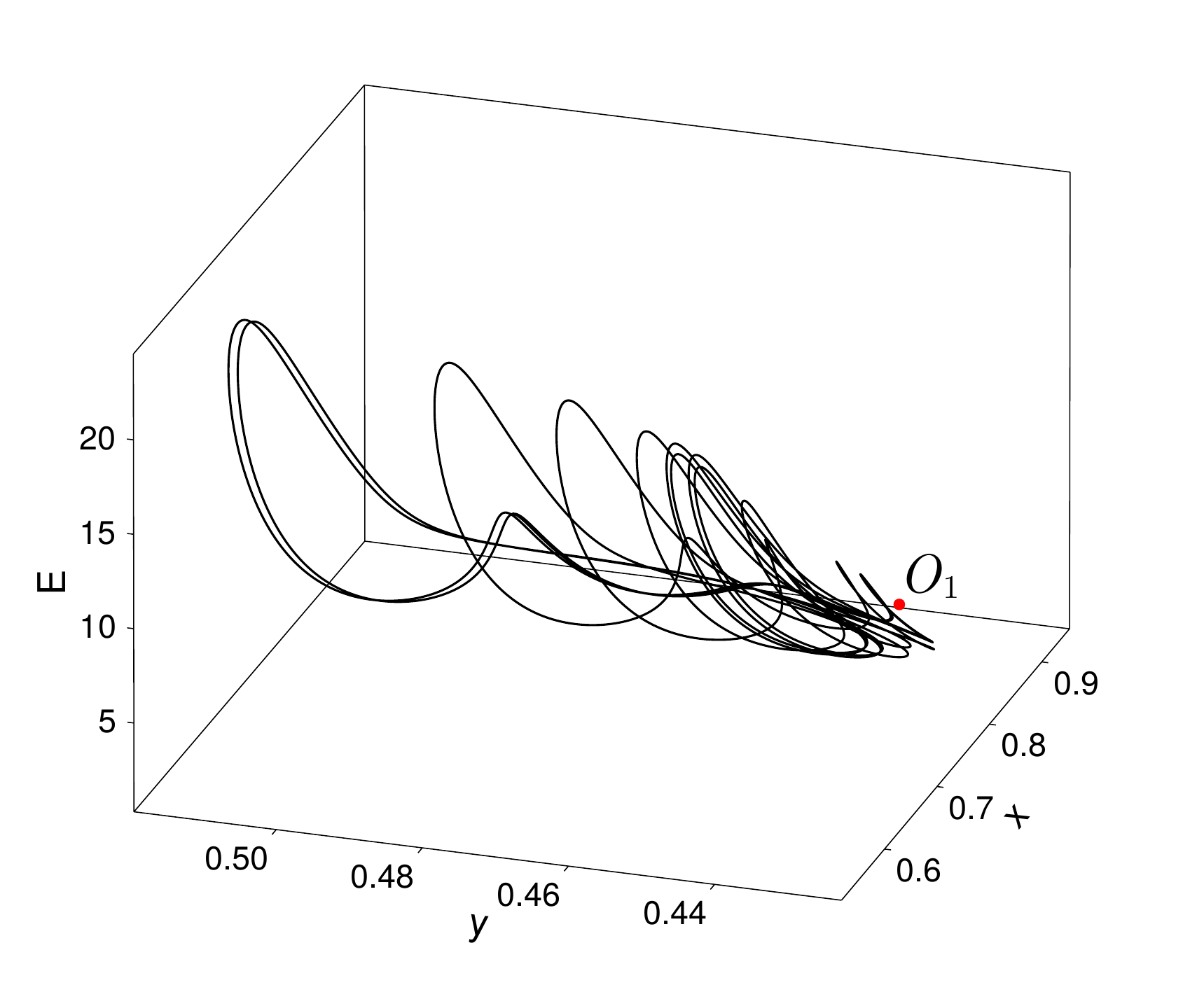}\\
    (b)\includegraphics[width= 14.0cm,  height = 4.0cm]{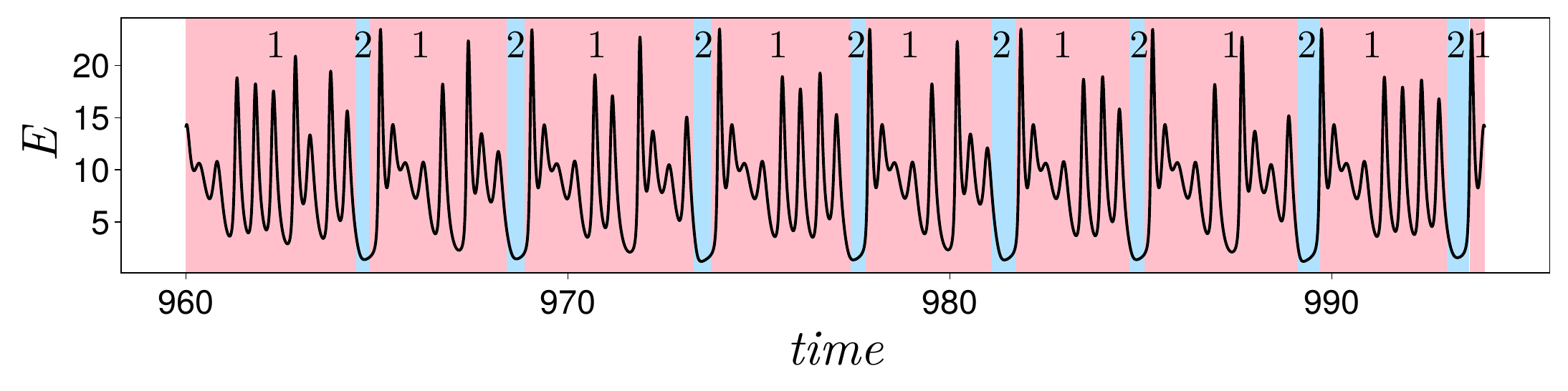}\\
	\caption{Chaotic non-homoclinic bursting activity (point A, $I_0 = -1.723$, $U_0 = 0.282$). (a) The phase portrait of the attractor. (b) Time series $E(t)$.}
	\label{fig:chaotic_non_hom}
\end{figure}

According to this remark, we can now divide all observed patterns of bursting population activity associated with spiral homoclinic attractors in system \eqref{eq:TM} into four types, depending on which attractor corresponds to them. Below we will give examples of such patterns, which are observed, in particular, at the values of the control parameters $I_0$ and $U_0$ corresponding to the selected points $A, B, C$ and $D$ on the MLE chart in fig. \ref{fig:LLE_map}. Here we follow  the classification introduced in \cite{bakhanova2018spiral}.

\subsection{Chaotic non-homoclinic bursting activity (point A)}

Such type of bursting is generated by non-homoclinic multi-funnel attractors, which appear in chaotic regions far from the homoclinic bifurcation curve $h_1$. Examples of phase portrait of this attractor can be seen in fig. \ref{fig:chaotic_non_hom}(a), while time series $E(t)$ are presented in fig. \ref{fig:chaotic_non_hom}(b). The main feature of this type of bursting population activity is a random number of fast oscillations (stage 1, red marker) that alternates with motions along the stable slow manifolds (stage 2, blue marker) of system \eqref{eq:TM}.

\subsection{Regular bursting activity (point B)}

\begin{figure}[tbh]
	\centering
    (a)\includegraphics[width = 9.0cm,  height = 9.0cm]{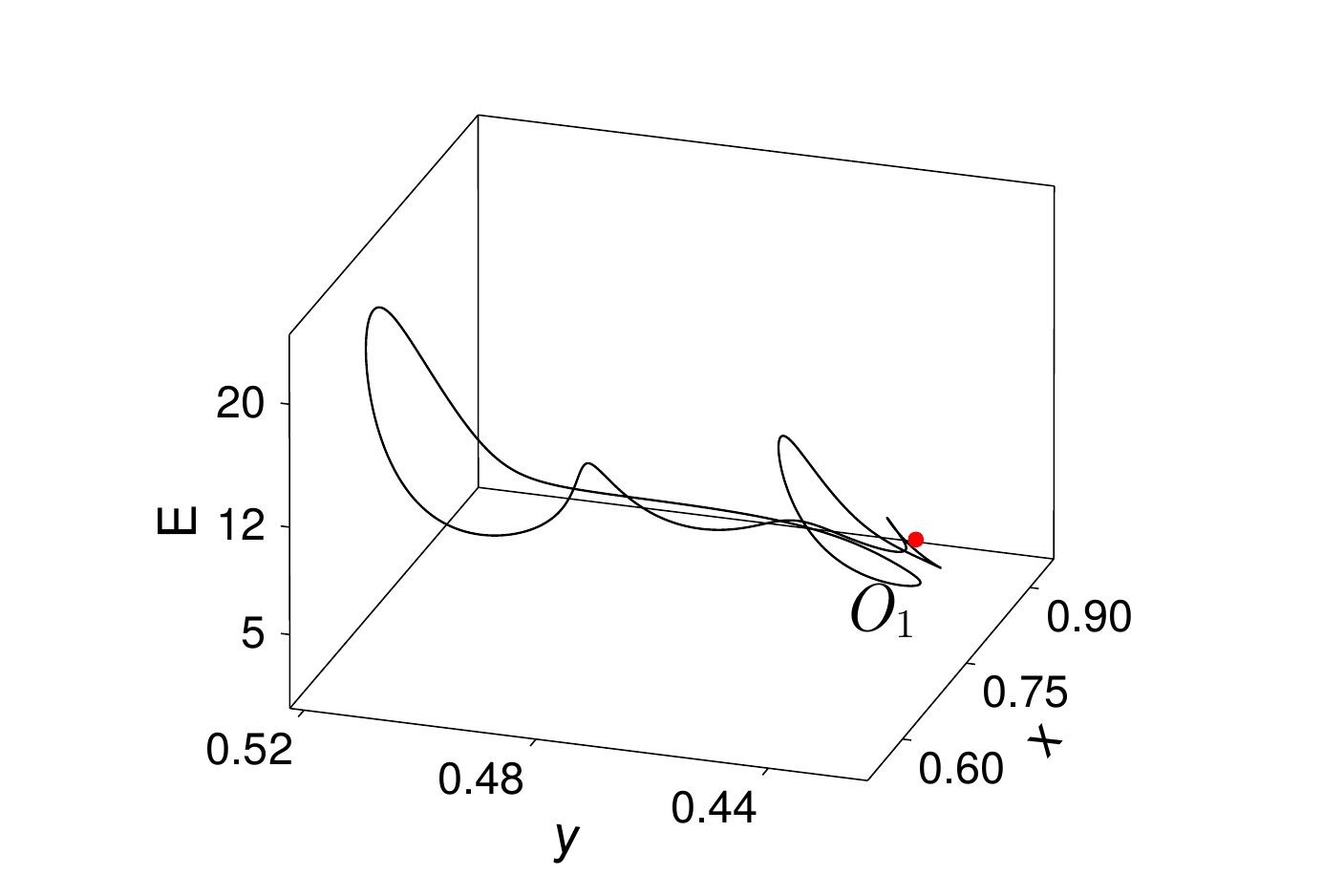}\\
    (b)\includegraphics[width= 14.0cm,  height = 4.0cm]{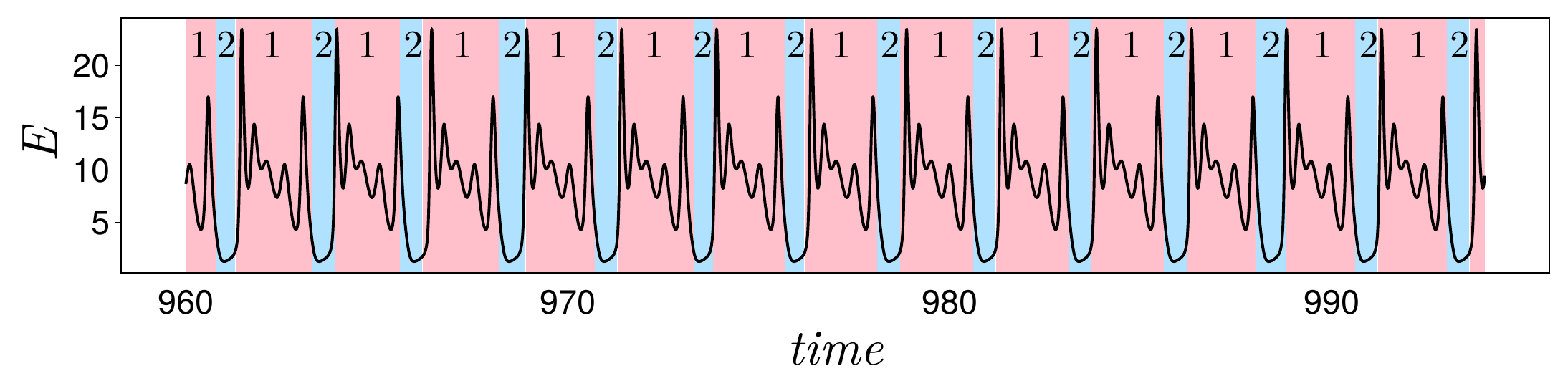}\\
	\caption{Regular non-homoclinic bursting activity (point B, $I_0 = -1.728$, $U_0 = 0.282$). (a) The phase portrait of the attractor. (b) Time series $E(t)$.}
	\label{fig:regular_burst}
\end{figure}

This type of bursting population activity is generated by the stable limit cycles, which appear in the stability windows of non-homoclinic multi-funnel attractors, see fig. \ref{fig:regular_burst}.  Such limit cycles repeat the behavior of trajectories on the corresponding attractors, and have the same number of turns. Comparing to the previous case (point A), the length of both stages is fixed.

\subsection{Chaotic homoclinic bursting activity (point C)}

\begin{figure}[tbh]
	\centering
    (a)\includegraphics[width=10.0cm,  height = 10.0cm]{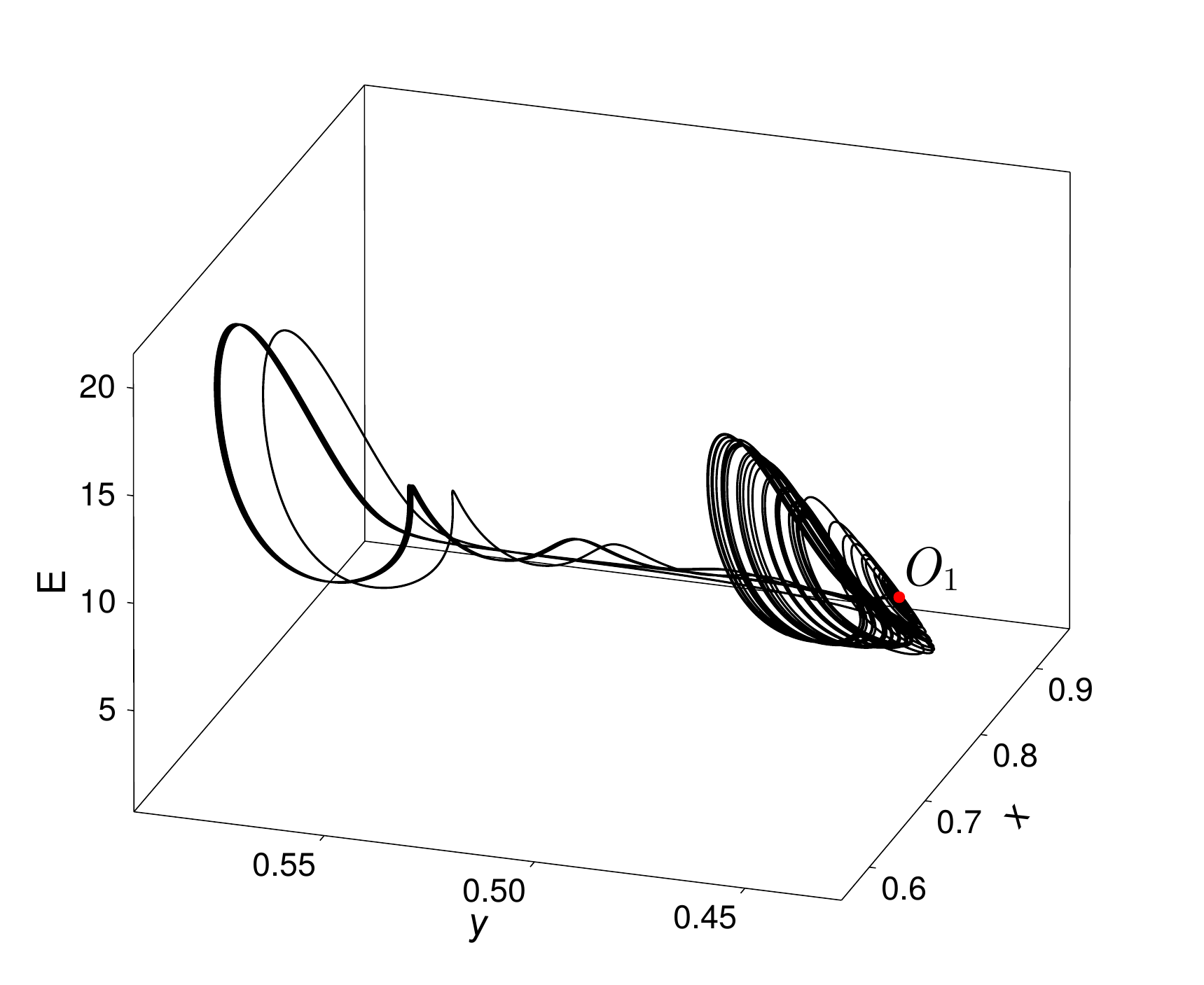}\\
    (b)\includegraphics[width= 14.0cm,  height = 4.0cm]{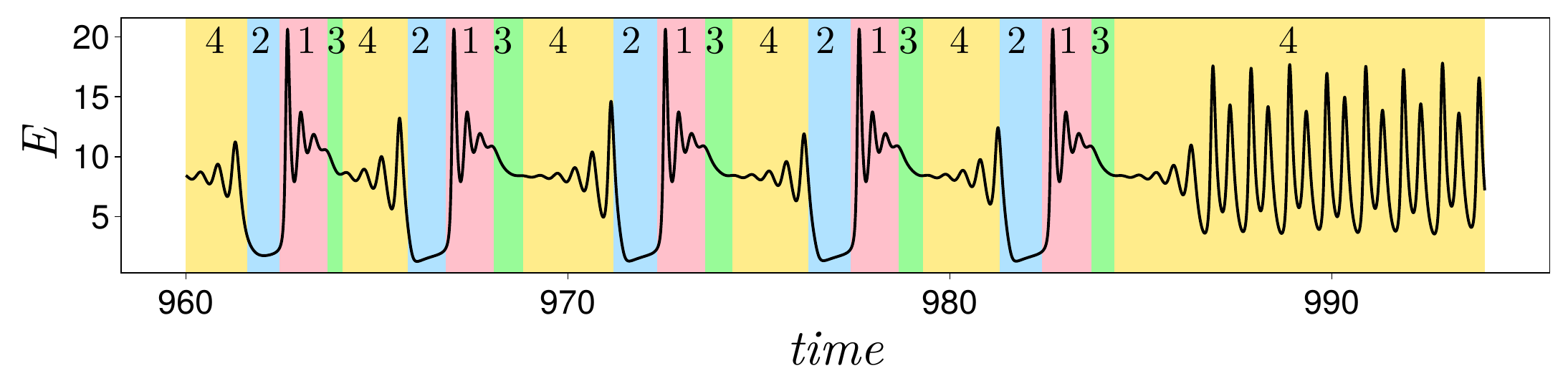}\\
	\caption{Chaotic homoclinic bursting activity (point C, $I_0 \approx -1.7071$, $U_0 \approx 0.26506$). (a) The phase portrait of the attractor. (b) Time series $E(t)$.}
	\label{fig:chaotic_homoclinic}
\end{figure}

This type of population bursting activity is associated with the co-existence of slow-fast behavior in system \eqref{eq:TM} with a drift near a homoclinic orbit  of a multi-funnel attractor. It appears in the chaotic regions along the homoclinic bifurcation curve $h_1$. As one can see in fig. \ref{fig:chaotic_homoclinic}, a random number of fast oscillations (stage 1, red marker) alternates with two types of slow motions: previously observed motions along the stable slow manifold of the fast subsystem (stage 2, blue marker) and small amplitude oscillations near a saddle-focus equilibrium (stage 3, green marker). Slow motions associated with trajectories passing near the saddle-focus $O_1$ appear when phase point returns to the neighbourhood of the saddle-focus after producing fast oscillations. The length of stage 3 depends on how close a trajectory passes near the saddle-focus $O_1$. %Thus, in contrast to chaotic non-homoclinic bursting, time interval between bursts can be unbounded. 
After the slow motions at stage 3 phase trajectory oscillates near the saddle-focus (stage 4, yellow marker) and then goes to the stable slow manifold (stage 2, blue marker).

\subsection{Regular near homoclinic bursting activity (point D)}

\begin{figure}[tbh]
	\centering
    (a)\includegraphics[width=9.0cm,  height = 9.0cm]{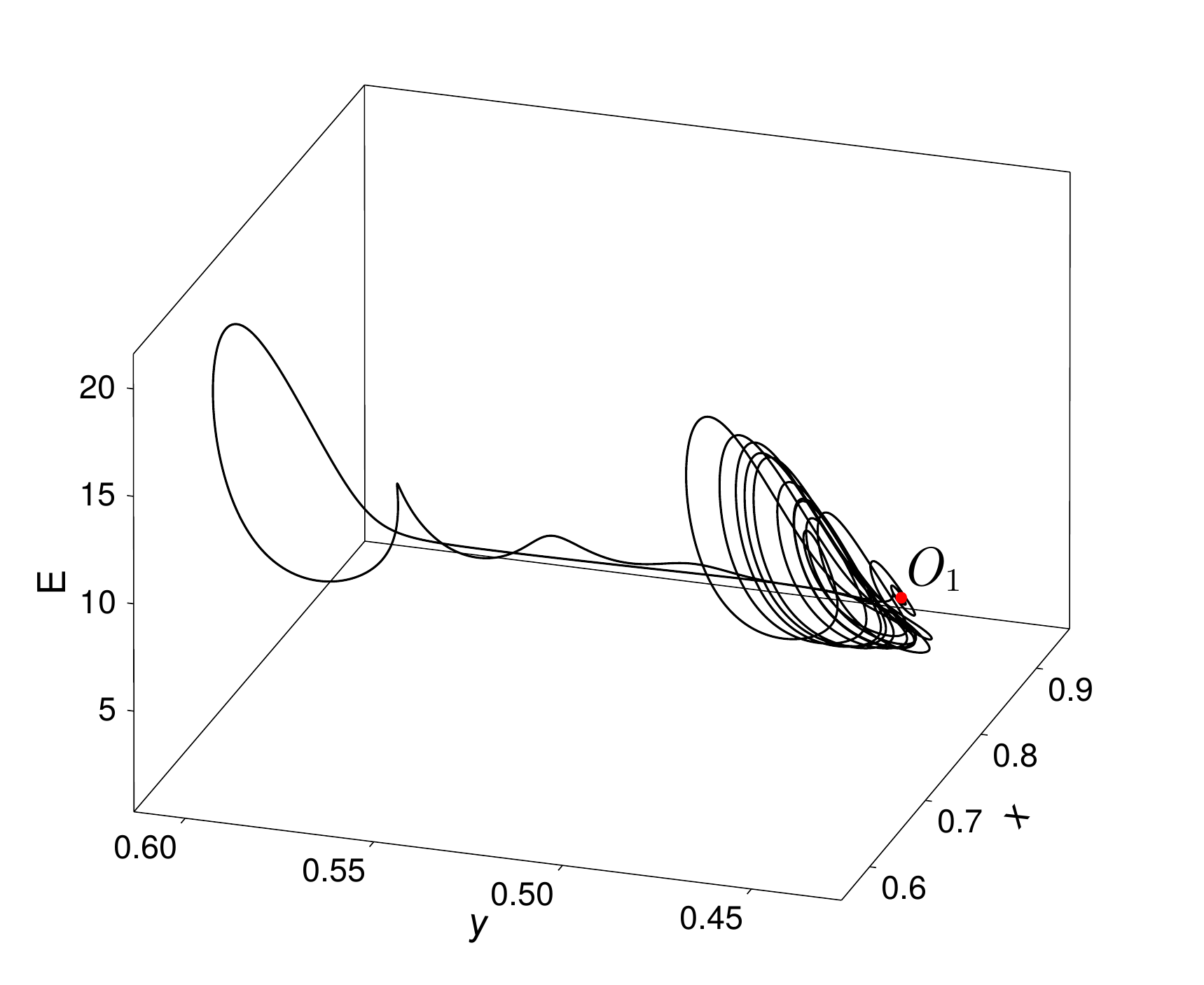}\\
    (b)\includegraphics[width= 14.0cm,  height = 4.0cm]{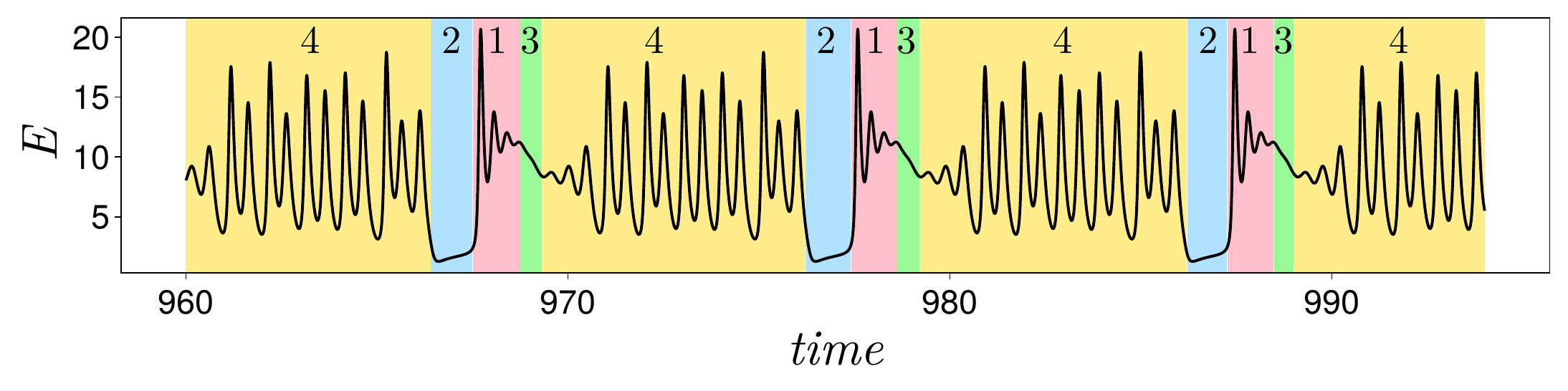}\\
	\caption{Regular near homoclinic bursting activity (point C, $I_0 = -1.714$, $U_0 = 0.2657$). (a) The phase portrait of the attractor. (b) Time series $E(t)$.}
	\label{fig:regularnearhomoclinic}
\end{figure}

This type of population bursting activity is associated with the multi-round limit cycles which appear close enough to the saddle-focus $O_1$, see \ref{fig:regularnearhomoclinic}(a). The motions along the corresponding part of these cycles are slow, see \ref{fig:regularnearhomoclinic}(b) (stage 3). Fast oscillations (stage 1) alternate with regular slow motions near the saddle-focus $O_1$ (stage 3) and with regular motions along the stable slow manifold of the fast subsystem (stage 2). The transition between stage 3 and stage 2 corresponds to the passing from a neighbourhood of the saddle-focus $O_1$ to the slow stable manifold. The lengths of all stages are fixed.

\begin{figure}[h!]
	\centering
    (a) \includegraphics[width=10.0cm,  height = 7.0cm]{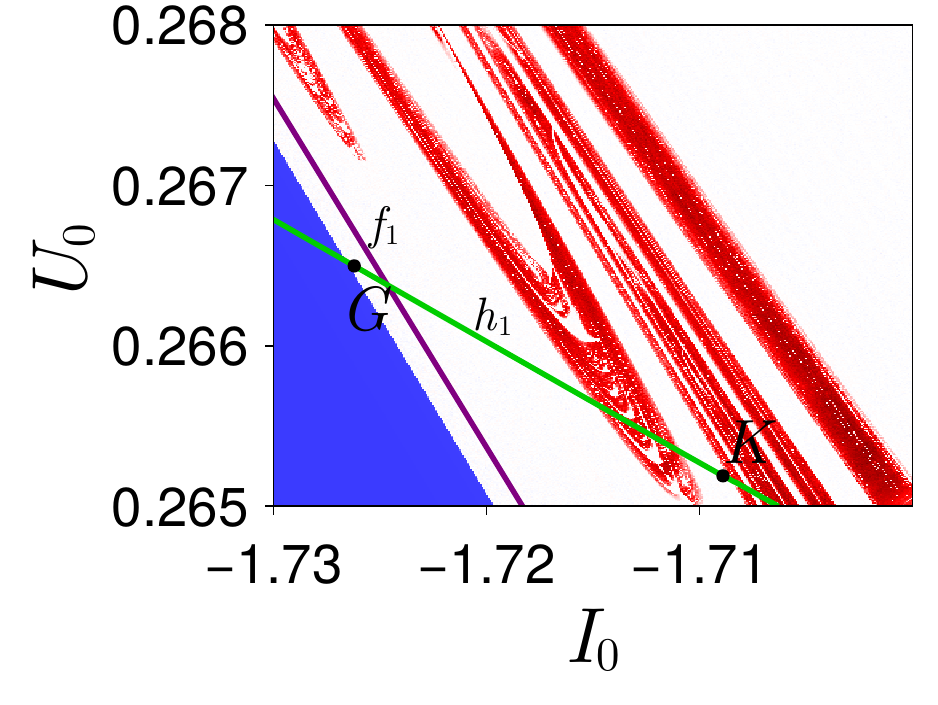}\\
    (b) \includegraphics[width=5.0cm,  height = 5.0cm]{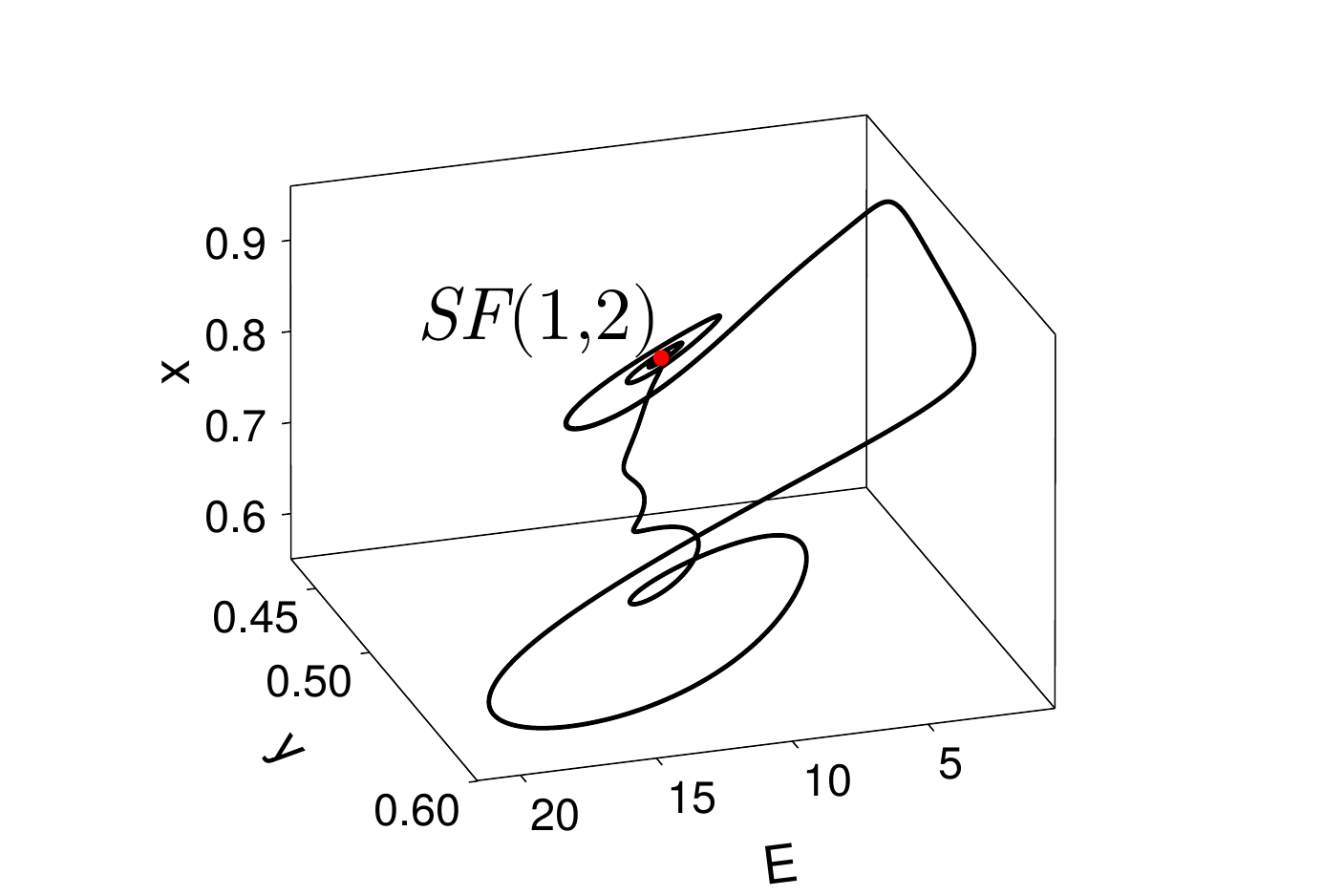}
    (c) \includegraphics[width= 10.0cm,  height = 4.0cm]{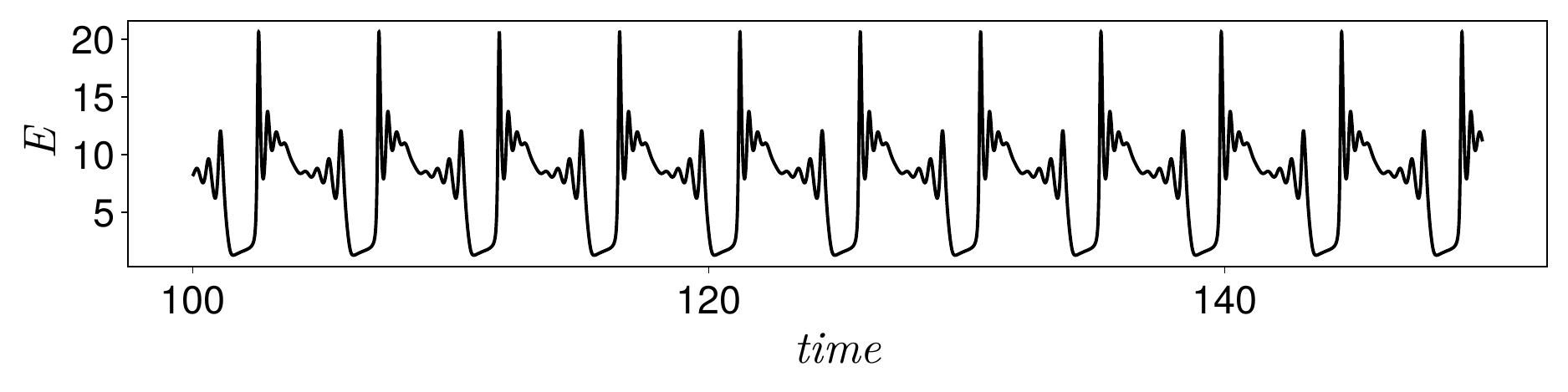}\\
    (d) \includegraphics[width=5.0cm,  height = 5.0cm]{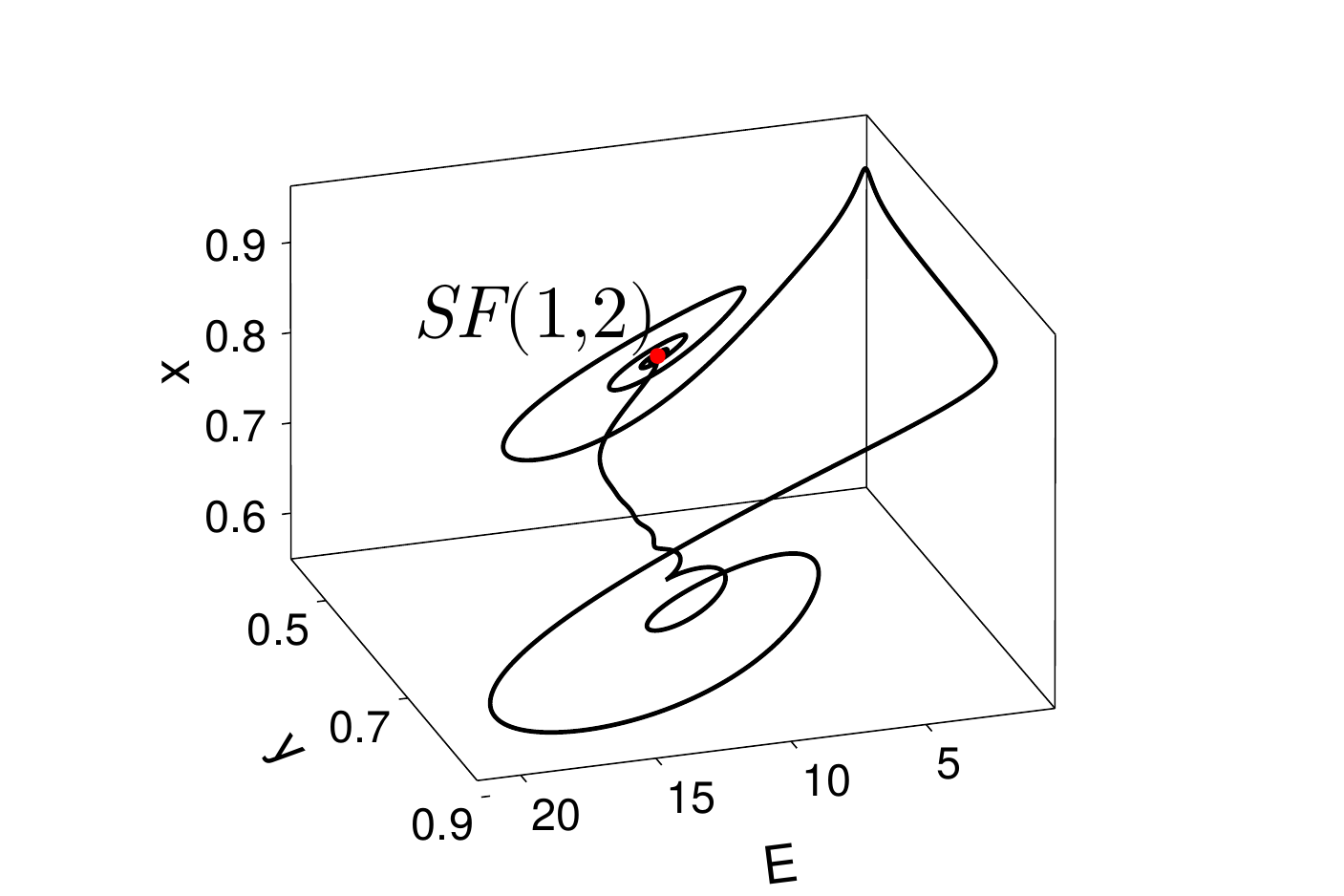}
    (e) \includegraphics[width= 10.0cm,  height = 4.0cm]{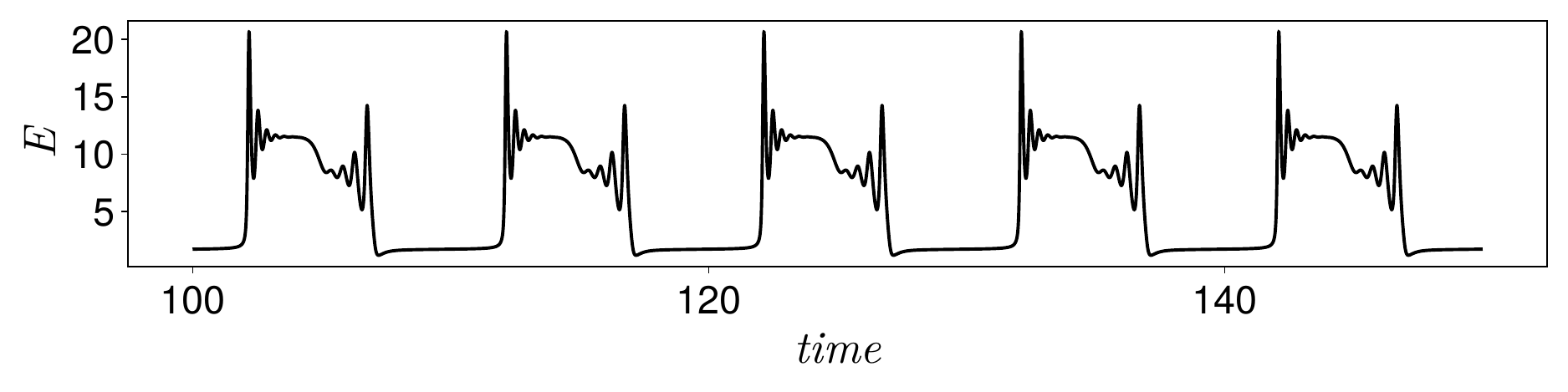}\\
	\caption{Examples of regular near homoclinic bursting activity. (a) Part of MLE chart in the vicinity of the intersection of $f_1$ fold bifurcation curve and $h_1$ homoclinic bifurcation curve. Points G and K mark combinations of governing parameters $I_0$ and $U_0$ that correspond to regular population bursting patterns with different length of quiescence stages. (b) The phase portrait and (c) time series $E(t)$ for $I_0 \approx -1.7088$, $U_0 \approx 0.2651$ (point K). (d) The phase portrait and (e) time series $E(t)$ for $I_0 \approx -1.7236$, $U_0 \approx 0.2663$ (point G).}
	\label{fig:intervalburst}
\end{figure}

As one can see in fig. \ref{fig:intervalburst}, the interburst interval increases while travelling along the homoclinic curve $h_1$ from right to left. At the left borderline of white region (fig. \ref{fig:intervalburst}(a)) saddle-node bifurcation on the homoclinic curve takes place. Before it the  called bottleneck occur -- a region in the phase space of system \eqref{eq:TM} where the phase point slows down along the trajectory, which is a sign of upcoming bifurcation. A part of attractor (stable slow manifold, stage 2) stretches towards a future equilibrium. A similar mechanism for increasing the interburst interval is observed outside the homoclinic bifurcation curve $h_1$ near the saddle-node bifurcation on the circle on the left border of the white region.

\section{Discussion}
\label{sec:discussion}

An important challenge in neuroscience is to understand how complex patterns of neural activity and brain functions are shaped by the collective dynamics of big populations of neurons. Most of the research on this problem has focused on large-scale numerical simulations \cite{brette2007simulation,izhikevich2008large,izhikevich2007dynamical}. An alternative is mean field theory, which provides insight into the macroscopic states of large neural networks in terms of collective neuronal activity or firing rates \cite{montbrio2015macroscopic}. This kind of analysis allows one to draw conclusions about a set of key macroscopic parameters for a  neuronal population. Knowledge of a set of such most important averaged characteristics of ensembles for describing their dynamics allows one to formulate phenomenological self-consistent models based on observations.

Our mean-field model of neuron-glial interactions is based on the well-known Wilson-Cowan model, which considers integral neuronal dynamics without reference to individual spikes. A distinctive feature of our model is the modeling of three components: the averaged population activity of neurons ($E$), the concentration of neuro- and gliotransmitters ($x$ and $y$, correspondingly), taking into account synaptic plasticity in the form of synaptic depression and astrocytic potentiation. The model allows one to study patterns of synchronized activity in large neuronal populations regardless of the specific number of neurons, local dynamics of neurons and glial cells, or the precise synaptic weights. In this approach all properties are averaged across the distribution of possible weights, and in the limit of infinite network size. All observed effects of neuron-like dynamics and neuron-glial interaction are general and are determined only by the presence of a feedback loop between the presynaptic neuron and the glial cell.

The proposed model allows us to reproduce and analyze various temporal patterns observed at the population level in big neuronal ensembles in biological experiments \cite{lebedeva2023effect,wagenaar2006persistent}. In particular, in \cite{wagenaar2006persistent} it was shown that neural activity observed in neuronal cultures using multielectrode recordings may include bursting and superbursting activity (microscopic level), which in our mean-field model correspond to oscillatory and bursting activity (macroscopic level). In this case, chaotic dynamics can be considered as one of the mechanisms of synchronization and switching between states of neural activity \cite{rabinovich1998role}.

\section{Conclusion}
\label{sec:conclusion}

In this paper we have studied scenarios for the birth of spiral homoclinic attractors in  system \eqref{eq:TM} that is a mean-field model of neuron-glial interaction. This model was proposed in \cite{barabash2023rhythmogenesis} and it describes the synchronous neuronal activity of a population of excitatory neurons under astrocytic modulation. We have shown that spiral attractors in system \eqref{eq:TM} arise according to the Shilnikov scenario, that leads to the appearance of a homoclinic attractor containing a homoclinic loop to the saddle-focus equilibrium with a two-dimensional unstable manifold. As a result, the system under study can generate several types of bursting population activity with different properties.

The obtained results can help to gain new insights into the nature of some specific patterns of activity that may arise in models of neuron-glial interactions. We believe that the obtained results are important not only from the point of view of nonlinear dynamics, but also from the point of view of neurodynamics and biology.

\begin{acknowledgments}
We are grateful to E.A. Grines, H.G.E. Meijer, R. Veltz, C. Rackauckas and G. Datseris for their helpful advice with numerical calculations. We also thank S.V. Gonchenko for careful reading. This study was financially supported by the Russian Science Foundation grant no.~19-72-10128.
\end{acknowledgments}

\section*{Data Availability Statement}

The data that support the findings of this study are available from the corresponding author upon reasonable request.

\appendix
\nocite{*}
\bibliography{aipsamp}
\end{document}